\documentclass[preprint,12pt]{elsarticle}

\usepackage{amssymb}

\usepackage[dvipsnames]{xcolor}

\usepackage{float}
\usepackage{caption}
\usepackage{subcaption}
\usepackage{graphicx}
\usepackage{amsmath}
\usepackage{chemfig}

\usepackage{makecell}
\usepackage{siunitx} 
\usepackage{booktabs} 

\usepackage{longtable}
\usepackage[colorinlistoftodos]{todonotes}
\usepackage{acro}%
\usepackage{hyperref}
\usepackage[utf8x]{inputenc}
\usepackage[draft]{changes}
\usepackage{gensymb}

\DeclareSIUnit{\cal}{cal}
\DeclareSIUnit\bar{bar}
\DeclareSIUnit\atm{atm}
\DeclareSIUnit\angstrom{\text {Å}}

\newcommand{\etal}{\textit{et al.~}}
\DeclareAcronym{AAD}{short=AAD, long=Average Absolute Deviation}
\DeclareAcronym{RMSD}{short=RMSD, long=Root Mean Square Deviation}
\DeclareAcronym{MD}{short=MD, long=Molecular Dynamics}
\DeclareAcronym{GC}{short=GC, long=Group Contribution}
\DeclareAcronym{HB}{short=HB, long=Hydrogen bond}
\DeclareAcronym{QM}{short=QM, long=Quantum Mechanics}
\DeclareAcronym{QC}{short=QC, long=Quantum chemical}
\DeclareAcronym{CPCM}{short=CPCM, long=Conductor-like polarizable continuum model}
\DeclareAcronym{IDAC}{short=IDAC, long=infinite dilution activity coefficient}
\DeclareAcronym{ML}{short=ML, long=Machine Learning}
\DeclareAcronym{D-MPNN}{short=D-MPNN, long=Directed-Message Passing Neural Network}
\DeclareAcronym{MPNN}{short=D-MPNN, long=Message Passing Neural Network}
\DeclareAcronym{GNN}{short=GNN, long=Graphical Neural Network}
\DeclareAcronym{RMSE}{short=RMSE, long=Root-Mean-Square Error}
\DeclareAcronym{QSPR}{short=QSPR, long=Quantitative Structure-Property Relationships}
\DeclareAcronym{vdW}{short=vdW, long=van der Waals}
\DeclareAcronym{API}{short=API, long=Active Pharmaceutical Ingredient}
\DeclareAcronym{COSMO-RS}{short=COSMO-RS, long=Conductor-like screening model for realistic solvation}
\DeclareAcronym{OF}{short=OF, long=Objective Function}
\date{}

\begin{document}\sloppy
\begin{frontmatter}

\title{Predicting solvation free energies for neutral molecules in any solvent with openCOSMO-RS}
\author[inst1]{Simon Müller}
\affiliation[inst1]{organization={Institute of Thermal Separation Processes, Hamburg University of Technology},
            city={Hamburg},
            postcode={21073},
            country={Germany}}

\author[inst2]{Thomas Nevolianis}
\affiliation[inst2]{organization={Institute of Technical Thermodynamics, RWTH Aachen University},
            city={Aachen},
            postcode={52062},
            country={Germany}}

\author[inst3]{Miquel Garcia-Ratés}
\author[inst3]{Christoph Riplinger}
\affiliation[inst3]{organization={FAccTs GmbH},
            city={Cologne},
            postcode={50677},
            country={Germany}}

\author[inst2]{Kai Leonhard}
\author[inst1]{Irina Smirnova}

\begin{abstract}
The accurate prediction of solvation free energies is critical for understanding various phenomena in the liquid phase, including reaction rates, equilibrium constants, activity coefficients, and partition coefficients. 
Despite extensive research, precise prediction of solvation free energies remains challenging.
In this study, we introduce openCOSMO-RS 24a, an improved version of the open-source COSMO-RS model, capable of predicting solvation free energies alongside other liquid-phase properties.
We parameterize openCOSMO-RS 24a using quantum chemical calculations from ORCA 6.0, leveraging a comprehensive dataset that includes solvation free energies, partition coefficients, and infinite dilution activity coefficients for various solutes and solvents at \SI{25}{\degreeCelsius}.
Additionally, we develop a Quantitative Structure-Property Relationships model to predict molar volumes of the solvents, an essential requirement for predicting solvation free energies from structure alone.
Our results show that openCOSMO-RS 24a achieves an average absolute deviation of \SI{0.45}{\kilo\cal\per\mol} for solvation free energies, \SI{0.76}{} for partition coefficients, and \SI{0.51}{} for infinite dilution activity coefficients, demonstrating improvements over the previous openCOSMO-RS 22 parameterization and comparable results to COSMOtherm 24 TZVP.
A new command line interface for openCOSMO-RS 24a was developed which allows easy acces to the solvation energy model directly from within ORCA 6.0.
This represents a significant advancement in the predictive modeling of solvation free energies and other solution-phase properties, providing researchers with a robust tool for applications in chemical and materials science.

\end{abstract}
\end{frontmatter}

\section{Introduction}

The accurate prediction of solvation free energies of solutes $\Delta G_{\mathrm{solv}}$ is crucial to understand phenomena occurring in the liquid phase.
From this quantity, one can determine varius thermodynamic and kinetic properties such as reaction rates, equilibrium  constants, activity coefficients, dissociation acidity constants, partition coefficients.
Consequently, solvation free energy plays a pivotal role in chemical reactions\cite{Mahalakshmi2008, Kroeger2017a, Heyden2018, Zhang2019, Nevolianis2023a} and design of materials with novel properties\cite{austin2017cosmo, austin2018cosmo, Gertig2019b, Zhou2020, Rasspe-Lange2022, Nevolianis2023d}.
Despite significant efforts over recent decades, precise prediction of $\Delta G_{\mathrm{solv}}$ remains a challenge.

In the last decade, various explicit\cite{Smith2005, Xi2022}, implicit\cite{Tomasi2005, Marenich2009/2, Klamt_95, klamt_refinement_1998, Eckert2010,stahn_extended_2023}, and data-driven\cite{Kang2007, Gilmer2017, Schweidtmann2020, Winter2022, SanchezMedina2022, Felton2022, Rittig2023} approaches have been used for solution phase property prediction.
Explicit approaches such as \ac{MD} are less common methods as they are quite computational time expensive as one needs to dissolve a solute in thousands of solvent molecules.
Implicit approaches are more common in solution phase property prediction since they are less computational demanding.
These approaches accurately predict the solvation free energies of neutral solutes with an uncertainty ranging from \SIrange{0.4}{1.1}{\kilo\cal\per\mol}\cite{Letcher2007, Marenich2009/2, Marenich2013, Qiu1997, Horta2016}.
Among others, the \ac{COSMO-RS} is a frequently used fully predictive implicit model with an uncertainty of \SIrange{0.40}{0.45}{\kilo\cal \per \mol} for predicting the solvation free energy of neutral solutes\cite{Letcher2007, Klamt2015, Vermeire2021}.
The basic principle of \ac{COSMO-RS} is based on the approximation of molecular interactions by the interactions of surface segments from the molecular cavities.
This makes the calculations less demanding than \ac{MD} calculations as the required input information only needs to be calculated once for each molecule from \ac{QM}.
Data-driven models have shown great potential in liquid phase property prediction mainly because many well established experimental databases are available.
For example, \ac{ML} methods have been quite promising in predicting the solvation free energies of neutral solutes\cite{Lim2021, Chung2022, FerrazCaetano2023}.
Vermeire \etal\cite{Vermeire2021}, trained and fine tuned a \ac{GNN} model for predicting solvation free energies of neutral molecules reporting an uncertainty of \SI{0.24}{\kilo\cal \per \mol}.
While these \ac{GNN} perform well on training datasets, their ability to generalize to new structures with different atoms and functional groups remains challenging. Empirical evidence suggests that their embeddings can generalize across different molecular spaces, but achieving robust out-of-distribution performance is still difficult\cite{Atz2021, Stark2021}. 

Recently, we published an open source version of the \ac{COSMO-RS} model, which we will call openCOSMO-RS 22\cite{Gerlach2022} in the following. This implementation of \ac{COSMO-RS} is the first open source version introducing additional descriptors besides the screening charge density. Having additional descriptors allowed the model to be modified for electrolytes with great success in the past\cite{gerlach_development_2018,muller_evaluation_2019,muller_calculation_2020,gonzalez_de_castilla_analogy_2021,gonzalez_de_castilla_analogy_2022,arrad_thermodynamic_2024}. For neutral molecules, openCOSMO-RS 22 performs quite well for predicting the \ac{IDAC} with a \ac{RMSD} of \SI{0.76}{} based on TURBOMOLE 6.6 parameterization and \SI{0.65}{} based on ORCA 5.0.3\cite{neese_orca_2012, neese_software_2018, Neese2020, neese_software_2022}.
Although openCOSMO-RS 22 was able to predict equilibrium properties between two or more solvents, it was not capable of predicting properties between gas and liquid phase such as the solvation free energy.

In this study, we perform a new parameterization of the model based on quantum chemical calculations from the software ORCA 6.0.
This will be called openCOSMO-RS 24a.
To do so, initially, we compile experimental data on solvation free energies, partition coefficients, and activity coefficients for a representative range of solutes and solvents.
Since the molar volume of the solvent is required to predict the solvation free energy, we develop a \ac{QSPR} model to predict the molar volume of the solvent at \SI{25}{\degreeCelsius} based on experimental data available in the literature\cite{Mathieu2016}.
We modified the openCOSMO-RS conformer workflow compared to that of our previous work\cite{Gerlach2022} by adding quantum chemical calculations of gas phase energies as these are needed to calculate the solvation free energies.
Leveraging the experimental data together with the \ac{QSPR} model, we parametrize openCOSMO-RS 24a for predicting the solvation free energies for a wide range of solutes and solvents.
During this work, we found that the Gaussian charge scheme, used within CPCM\cite{Barone1998, York1999} in ORCA\cite{garcia-rates_effect_2020} produced very small segments leading to unusually large screening charge densities. This was addressed by rejecting the addition of segments smaller than a specified threshold (\SI{0.01}{\angstrom\squared}) with minimum effect on the calculated energies. Furthermore, in ORCA, a Lagrangian-based algorithm is used to calculate the outlying charge correction\cite{Pye1999}.
Although this is not a treatment as advanced as the method proposed by Klamt \cite{klamt_treatment_1996} as it neglects the spacial distribution of the outlying charge, it should be enough for the neutral molecules tested. In the future a more thorough analysis of this is planned as it becomes especially important for anions.
Finally, we report the performance of openCOSMO-RS 24a, which as of now can directly be used within the ORCA 6.0 software as additional solvation model enabling the user to access a variety of liquid phase properties which previously was not possible.

\section{Methods}
\subsection{openCOSMO-RS}
The theory of openCOSMO-RS has been discussed in previous studies\cite{Kroeger2020, Gerlach2022} and only the equation related to solvation free energy is briefly summarized here.
The solvation free energy can then be calculated similarly to Klamt \etal\cite{Klamt1995, klamt_refinement_1998} from

\begin{equation}
\begin{aligned}
 \Delta G_{\mathrm{solv}}
      = E_{\mathrm{diel}}
      +RT\ln\gamma ^{\infty}
      -\sum\limits_{\mathrm{\alpha} }{\tau_{\mathrm{\alpha} }A_{\mathrm{\alpha} }}
      -\omega_{\mathrm{ring}}n_{\mathrm{ring}}
      -RT\ln\frac{\nu_{\mathrm{IG}}}{\nu_{\mathrm{liquid}}}
      -\eta
      \end{aligned}
      \label{Eq:dGsolv}
\end{equation}

The term $E_{\mathrm{diel}}$ represents the dielectric energy, which is the energy involved in transferring the solute from the gas phase to an ideal conductor.
The second term refers to the chemical potential at infinite dilution in the liquid phase, using the ideal conductor as the reference state and it is directly predicted by openCOSMO-RS.
The third term encompasses the energy required for cavity formation and includes a van der Waals-like contribution to the solvation free energy, calculated by summing the product of each atom's area $A_{\mathrm{\alpha }}$ on the solute molecule and a factor $\tau _{\mathrm{\alpha} }$ that depends on the atomic number.
The fourth term provides a correction for molecules containing rings, determined by multiplying a general parameter $\omega_ {\mathrm{ring}}$ by the number of rings $n_{\mathrm{ring}}$ in the solute structure.
The fifth term accounts for the change in units of the reference states from mole fraction (units of the calculation) to molar concentration (units of the experimental data) with $\nu _{\mathrm{IG}}$ and $\nu _{\mathrm{liquid}}$ representing the molar volumes of the ideal gas and the liquid phase, respectively.
The final term $\eta _{\mathrm{type}}$ is an adjustable parameter.

\subsection{Computational details}
In the following, we describe the openCOSMO-RS 24a conformer workflow for searching and calculating all necessary input data  for the gas and \ac{CPCM} phase.
This is the updated overview of the pipeline also available on github\cite{tuhh_tvt_github}:
\begin{itemize}
    \item Gas phase calculations
    \begin{itemize}
        \item[$-$] Molecular mechanics-based conformer generation using RDKit\cite{ebejer_freely_2012, noauthor_rdkit_nodate}.
        \item[$-$] Filter conformers by an energy window of \SI{6}{\kilo\cal\per\mol}.
        \item[$-$] Cluster conformers by an \ac{RMSD} window of \SI{1} and save these for \ac{CPCM}\cite{garcia-rates_effect_2020} calculations.
        \item[$-$] Geometry optimizations at DFT/BP86/def2-TZVP(-f)\cite{Becke1988, Perdew1986, weigend_balanced_2005} level using ORCA.
        \item[$-$] Single point energy calculation using DFT/BP86/def2\nobreakdash-TZVPD level in ORCA for the conformer with the lowest energy.
    \end{itemize}
    \item \ac{CPCM} calculations
    \begin{itemize}
        \item[$-$] Optimize geometries in water using ALPB \cite{ehlert_robust_2021} with GFN2-xTB \cite{Bannwarth2019} calculations from within ORCA, starting from saved conformers.
        \item[$-$] Filter confomers by an energy window of \SI{6}{\kilo\cal\per\mol}.
        \item[$-$] Cluster conformers by an \ac{RMSD} window of \SI{1} and select the conformers with the three lowest energies.
        \item[$-$] \ac{CPCM} geometry optimizations at the DFT/BP86/def2-TZVP(-f) level in ORCA.
        \item[$-$] Filter confomers by an energy window of \SI{6}{\kilo\cal\per\mol}.
        \item[$-$] Cluster conformers by an \ac{RMSD} window of \SI{1} and select the conformer with the lowest energy.
        \item[$-$] \ac{CPCM} geometry optimizations of DFT/BP86/def2\nobreakdash-TZVP level in ORCA.
        \item[$-$] \ac{CPCM} single point energy calculation using DFT/BP86/def2\nobreakdash-TZVPD level in ORCA.
    \end{itemize}
\end{itemize}

To search a large parameter space, the global solver differential evolution as implemented in SciPy\cite{2020SciPy-NMeth} is used.
Similar to our previous studies\cite{Kroeger2020, Gerlach2022}, the following objective function is minimized for all optimizations:

\begin{equation} \label{eqn:ofidac}
   \mathrm{OF} = \frac{1}{N_{p}}\sum_i{\left (Y^{\mathrm{calc}} - Y^{\mathrm{exp}} \right)^2}
\end{equation}

The average absolute deviation is calculated from:
\begin{equation} \label{eqn:aadidac}
   \mathrm{AAD}_{\mathrm{Y}} = \frac{1}{N_{\mathrm{p}}} \sum_i{\left| Y^{\mathrm{calc}} - Y^{\mathrm{exp}} \right|}
\end{equation}
whereby $\mathrm{Y}$ is either $\ln{\gamma_i^{\infty}}$,  $\ln{K}$ or $\Delta G_{\mathrm{solv}}$.

\subsection{Dataset overview}
The dataset used in this work is comprised of three data types at \SI{25}{\degreeCelsius}: (i) infinite dilution activity coefficients, (ii) partition coefficients and (iii) solvation free energies.
The 800+ infinite dilution activity coefficients are taken from Parcher \etal\cite{parcher_specific_1975}, Voutsas \& Tassios\cite{voutsas_prediction_1996}, Kontogeorgis \etal\cite{kontogeorgis_improved_1997}, Kato \etal\cite{kato_infinite_2002} and He \& Zhong\cite{he_qspr_2003}.
The partition coefficients for the following solvent combinations: octanol + water, benzene + water, hexane + water, and diethyl ether + water are collected by Klamt \etal\cite{klamt_refinement_1998}.
The 2000+ solvation free energies are taken from Marenich \etal\cite{Marenich2007}.
Xylene is excluded from the calculations as it is a mixture of constitutional isomers.
Additionally, values for water as a solute in all three data types were excluded due to their known prediction issues when solvated in non-polar solvents within the \ac{COSMO-RS} framework without further model improvements\cite{klamt_refinement_1998,klamt_prediction_2003}. Even for molecular simulations treating mixtures of water and alkanes over the complete concentration range is challenging for most models. \cite{ballal_isolating_2014, ballal_erratum_2016, asthagiri_electrostatic_2017}

To calculate the solvation free energy, the molar volume of the pure solvent is required (see Equation \ref{Eq:dGsolv}).
Thus, we develop a \ac{QSPR} model in this study to predict the molar volume of the solvent at \SI{25}{\degreeCelsius} based on experimental data available from Mathieu \& Bouteloup\cite{Mathieu2016}.
The complete dataset is cleaned and normalized: isotopes and explicit hydrogens are deleted, duplicates are merged, and only the first value in the original data for each component is retained.


\section{Results and Discussion}

\subsection{Predictive QSPR model for molar volumes of the solvent at \SI{25}{\degreeCelsius}}

\begin{figure}[h]
\centering
\includegraphics[width=0.8\textwidth]{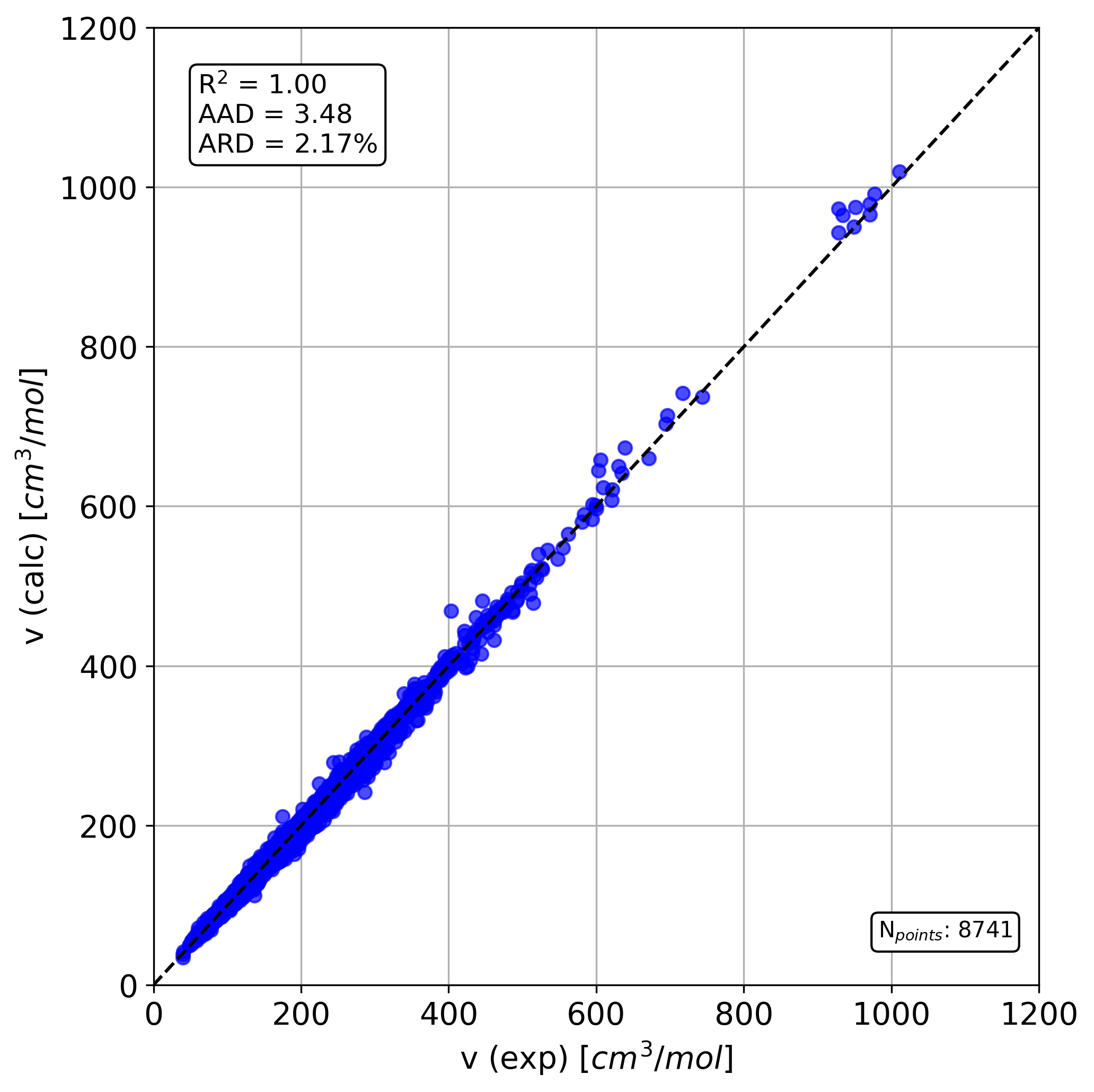}
\caption{Parity plot for the prediction of molar volume of the solvent model.}
\label{fig:mol_vol}
\end{figure}

To enable the fully predictive calculation of solvation energies, a model is developed to predict the only quantity not calculated within openCOSMO-RS; the molar volume of the pure solvent. 
The model is based on a linear combination of descriptors, represented by the following equation:

\begin{equation} \label{eqn:molar_volume}
\begin{aligned}
    v_{pure} = 0.6977 A_{\mathrm{CPCM}} - 0.3161 M_2 + 0.03244 M_4\\ + 0.9431 n_{\mathrm{atoms}} + 8.113 n_{\mathrm{Si,atoms}} - 0.07067
\end{aligned}
\end{equation}

where $A_{\ac{CPCM}}$ is the area of the surface segments on the cavity of the solute, $M_{\mathrm{i}}$ are the respective sigma moments, $n_{\mathrm{atoms}}$ is the number of atoms and $n_{\mathrm{Si,atoms}}$ is the number of silicon atoms in the molecule.
All descriptor combinations were systematically evaluated. Notably, $A_{\ac{CPCM}}$ offers a more effective representation of the molar volume of the solvent compared to the volume of the cavity while utilizing fewer descriptors.
The significant effect of the number of silicon atoms on the model's accuracy might suggest that the silicon radius might not be optimal.
Figure \ref{fig:mol_vol} shows the predicted molar volumes of the solvents using the \ac{QSPR} molar volume of the solvent model against the experimental molar volumes of the solvents at \SI{25}{\degreeCelsius} based on experimental data described more in detail in the previous section. 
Overall, the predicted molar volumes of the solvents agree well with the experimental ones with \ac{AAD} of \SI{3.48}{\centi\meter\cubed \per \mole} and R$^2$ of 0.995.
Mathieu and Bouteloup report a model for predicting the standard density with an average relative error $<$1.7\%. For density prediction, our \ac{QSPR} model achieves a relative error of 2.2\%, which is an accuracy similar to that of other group contribution methods\cite{Elbro1991, Ihmels2003, Hukkerikar2012}.

\subsection{Parametrization}
All non-fixed parameters in Table \ref{tbl:parameterizations} are simultaneously adjusted using the differential evolution algorithm implemented in SciPy\cite{2020SciPy-NMeth}.
Following the approach used in openCOSMO-RS 22\cite{Gerlach2022}, we incorporate the improved misfit term, which includes the additional descriptor $\sigma^{\perp}$ to recover some of the lost 3D information.
All data are included in the regression of the parameters.

\begin{table}[h]
  \centering
  \caption{Parameterization of openCOSMO-RS 24a based with gas and \ac{CPCM} geometry optimizations at DFT/BP86/def2-TZVP level and gas and \ac{CPCM} single point calculations at DFT/BP86/def2-TZVPD level in ORCA 6.0. [*] denotes the parameter was fixed.}
  \label{tbl:parameterizations}
  \begin{tabular}{l c l c}
    \toprule
    \textbf{Parameter} & \textbf{Value} & \textbf{Parameter} & \textbf{Value} \\
    \midrule
    $r_{av}^* \; [\AA]$ & 0.5 & $\tau_1 \left[\frac{kJ}{mol \cdot \AA^2}\right]$ & 0.123 \\
    $a_{eff} \; [\AA^2]$ & 5.925 & $\tau_6 \left[\frac{kJ}{mol \cdot \AA^2}\right]$ & 0.096 \\
    $\alpha_{mf} \left[\frac{kJ \cdot \AA^2}{mol \cdot e^2}\right]$ & 7281 & $\tau_7 \left[\frac{kJ}{mol \cdot \AA^2}\right]$ & 0.003 \\
    $f_{corr}^* [-]$ & 2.4 & $\tau_8 \left[\frac{kJ}{mol \cdot \AA^2}\right]$ & 0.015 \\
    $c_{hb} \left[\frac{kJ \cdot \AA^2}{mol \cdot e^2}\right]$ & 43327 & $\tau_9 \left[\frac{kJ}{mol \cdot \AA^2}\right]$ & 0.023 \\
    $\sigma_{hb} \; [e/\AA^2]$ & 0.00961 & $\tau_{17} \left[\frac{kJ}{mol \cdot \AA^2}\right]$ & 0.143 \\
    $A_{std} \; [\AA^2]$ & 41.624 & $\tau_{53} \left[\frac{kJ}{mol \cdot \AA^2}\right]$ & 0.891 \\
    $\eta \left[\frac{kJ}{mol}\right]$ & -18.61 & $\tau_{14} \left[\frac{kJ}{mol \cdot \AA^2}\right]$ & 0.018 \\
    $\omega_{ring} \left[\frac{kJ}{mol}\right]$ & 1.100 & $\tau_{15} \left[\frac{kJ}{mol \cdot \AA^2}\right]$ & 0.015\\
    & & $\tau_{16} \left[\frac{kJ}{mol \cdot \AA^2}\right]$ & 0.146 \\
    \bottomrule
  \end{tabular}
\end{table}

\subsection{Model performance}
\begin{figure}[h]
\centering
\includegraphics[width=0.8\textwidth]{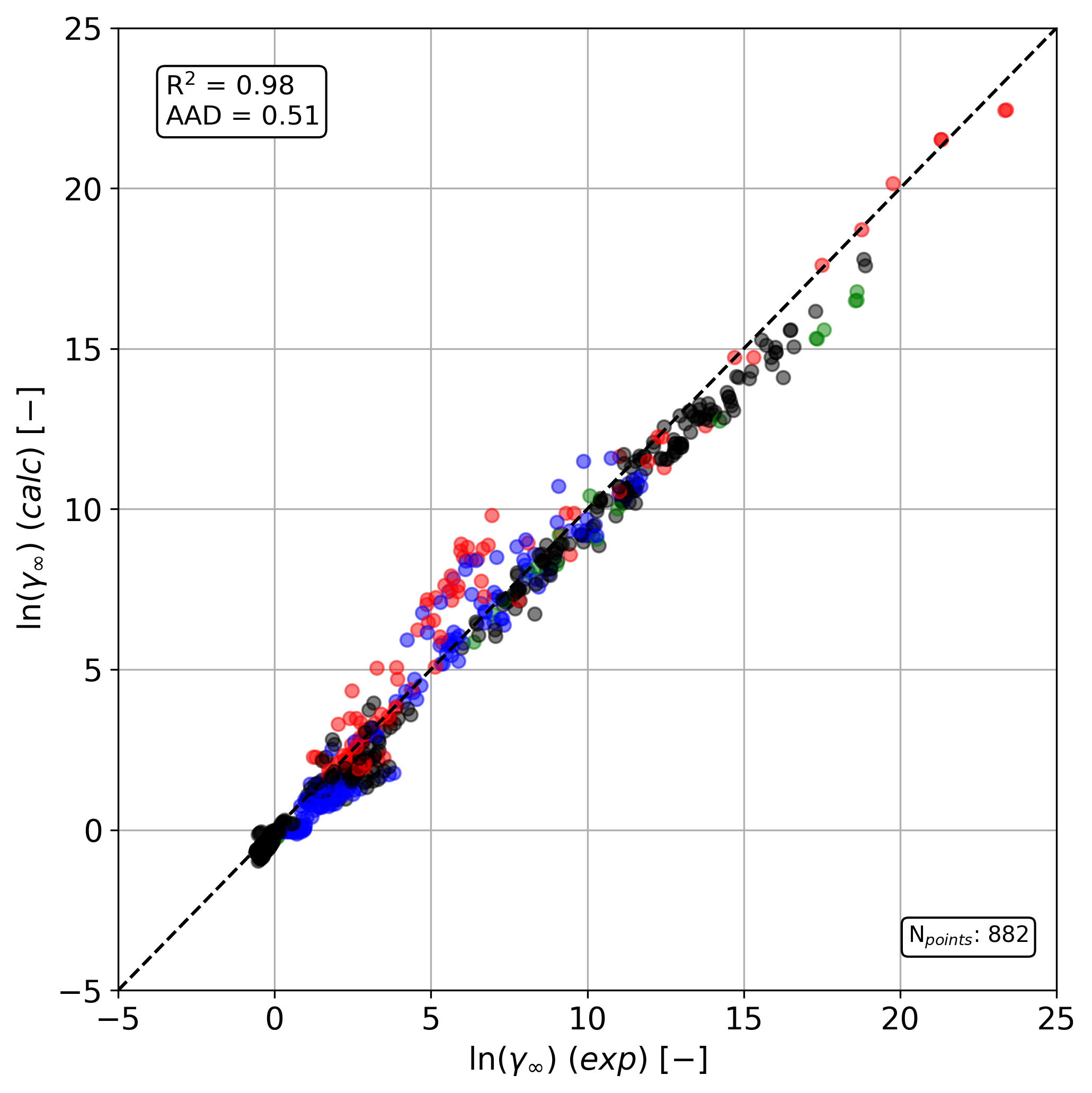}
\caption{Parity plot for infinite dilution activity coefficients calculated with openCOSMO-RS 24a. Colors represent different solute types: ({\large{$\bullet$}}) non-HB, ({\large\textcolor{blue}{$\bullet$}}) HB acceptors, ({\large\textcolor{OliveGreen}{$\bullet$}}) HB donors and ({\large\textcolor{red}{$\bullet$}}) HB donors/acceptors.}
\label{fig:gamma}
\end{figure}

\begin{figure}[h]
\centering
\includegraphics[width=0.8\textwidth]{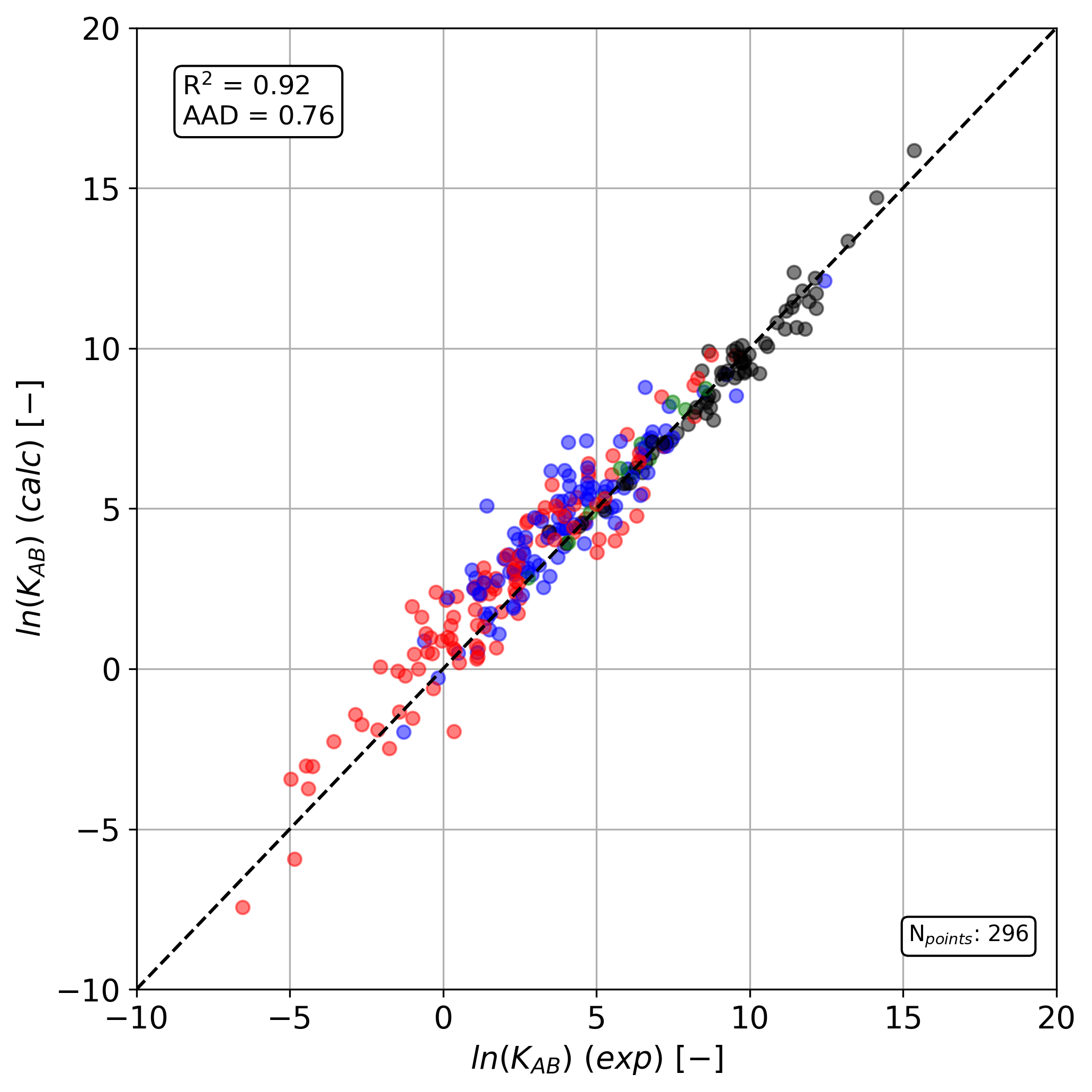}
\caption{Parity plot for partition coefficients at \SI{25}{\degreeCelsius} calculated with openCOSMO-RS 24a whereby AB = [octanol/water, benzene/water, hexane/water, diethyl ether/water]. Colors represent different solute types: ({\large{$\bullet$}}) non-HB, ({\large\textcolor{blue}{$\bullet$}}) HB acceptors, ({\large\textcolor{OliveGreen}{$\bullet$}}) HB donors and ({\large\textcolor{red}{$\bullet$}}) HB donors/acceptors.}
\label{fig:logp}
\end{figure}

\begin{figure}[h]
\centering
\includegraphics[width=0.8\textwidth]{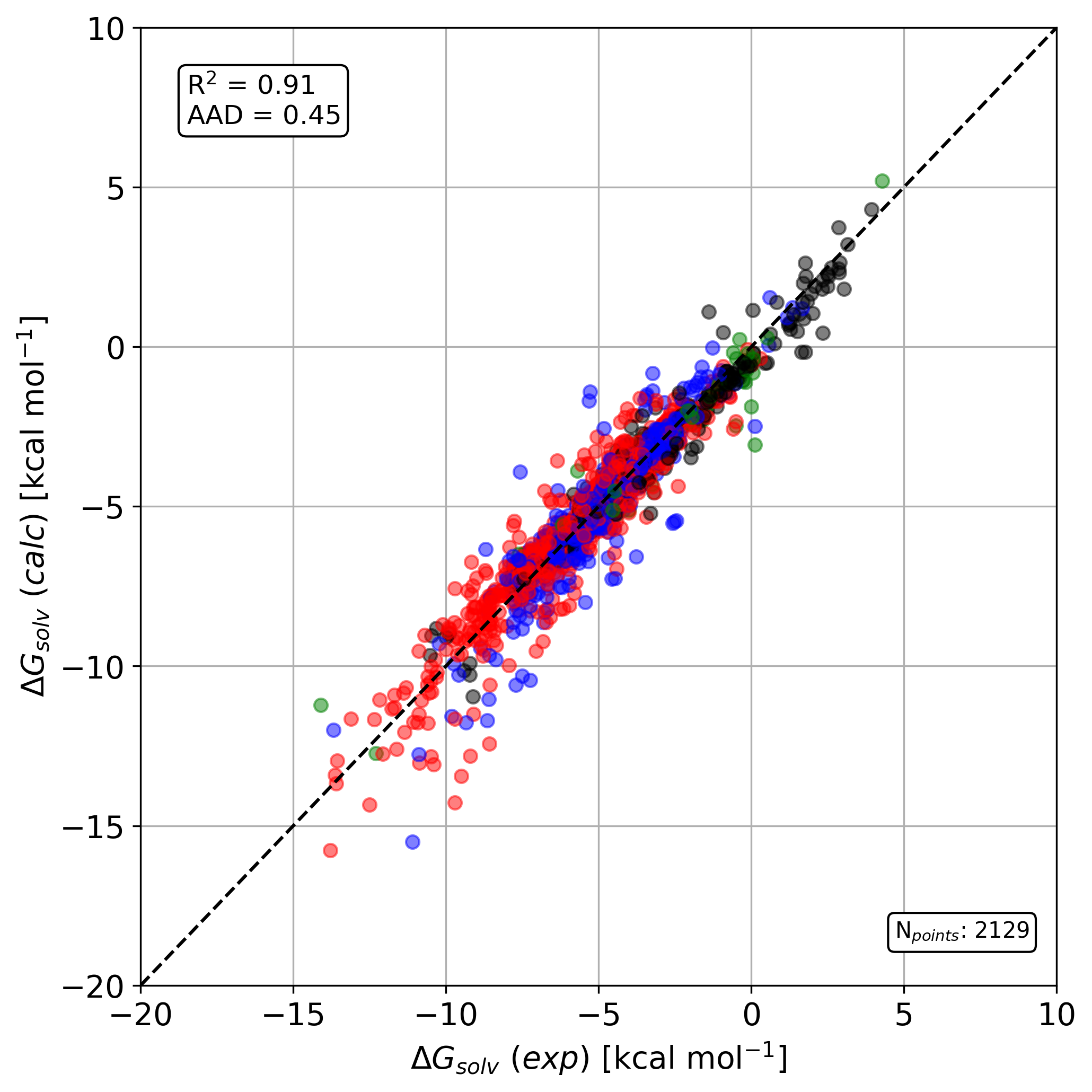}
\caption{Parity plot for solvation free energies at \SI{25}{\degreeCelsius} calculated with openCOSMO-RS 24a. Colors represent different solute types: ({\large{$\bullet$}}) non-HB, ({\large\textcolor{blue}{$\bullet$}}) HB acceptors, ({\large\textcolor{OliveGreen}{$\bullet$}}) HB donors and ({\large\textcolor{red}{$\bullet$}}) HB donors/acceptors.}
\label{fig:dgsolv}
\end{figure}

In this work, infinite dilution activity coefficients, partition coefficients, and solvation free energies, all at \SI{25}{\degreeCelsius} are used to parameterize openCOSMO-RS 24a.
Table \ref{tbl:comparison} provides an overview of all calculations for openCOSMO-RS 24a, COSMOtherm 24 TZVP and COSMOtherm 24 FINE, calculated with the lowest energy conformer.
Figures \ref{fig:gamma}, \ref{fig:logp}, and  \ref{fig:dgsolv} show the predicted values obtained from openCOSMO-RS 24a against the experimental values compiled from the literature for the activity coefficients, solvation free energies, and partition coefficients, respectively.
In all Figures, black represents solutes without hydrogen bonds, red represents solutes that are hydrogen bond donors, and blue represents solutes that are hydrogen bond acceptors.
Whether or not a solute is considered hydrogen bonding depends on the existence of area having a screening charge density larger than the threshold hydrogen bonding parameter $\sigma_{\mathrm{HB}}$.

For the infinite dilution activity coefficients, a total of 882 data points are used (see Figure \ref{fig:gamma}).
openCOSMO-RS 24a achieves an \ac{AAD} of \SI{0.51}{} and R$^2$ of \SI{0.98}{}, showing an improvement compared to our previous work openCOSMO-RS 22\cite{Gerlach2022}, which had an \ac{AAD} of \SI{0.65}.
Overall, it can be observed, that overall, for the less polar solutes, the infinite dilution activity coefficients are slightly underestimated while for the more polar ones are somewhat overestimated.
The infinite dilution activity coefficient represents the energy required for transferring one molecule from pure component to being infinitely dilute in a second solvent.
Hence, this systematic shift in model performance based on solute polarity might be either due to the overestimation of attractive hydrogen bonding for more polar molecules in the reference state (i.e. pure solute) or due to an overestimation of the repulsive misfit energy at infinite dilution in the solvent.
This will be investigated further in future work.

For the partition coefficients, the dataset includes 296 data points (see Figure \ref{fig:logp}). 
The openCOSMO-RS 24a models achieves an \ac{AAD} of \SI{0.76}{} and R$^2$ of \SI{0.92}{}, indicitating good agreement between the calculated and the experimental data.
Similar to \ac{IDAC}, the model tends to overestimate partition coefficients for more polar solutes and slightly underestimate them for less polar ones.
The partition coefficient measures the energy required to transfer a solute at infinite dilution from water to another solvent, representing the relative interaction energy of the solute with the other solvent compared to water.
The data suggests that the greater the polarity difference is between the solute and the other solvent, the larger the deviation in calculated values by openCOSMO-RS 24a.
As mentioned earlier, systems with water as a solute are excluded from the dataset because the usual COSMO-RS theory struggles to handle water at infinite dilution in very non-polar solvents~\cite{klamt_refinement_1998,klamt_prediction_2003}.
This issue appears to extend to other polar molecules in non-polar solvents, though less pronounced than with water.
This suggests a general systematic issue that could potentially be addressed to improve the model.

For the solvation free energies, the dataset contains 2129 data points (see Figure \ref{fig:dgsolv}).
The openCOSMO-RS 24a model achieves an \ac{AAD} of \SI{0.45}{\kilo\cal \per \mol} and R$^2$ of \SI{0.91}{}, showing a strong agreement between the calculated and experimental values, which is impressive considering that the molecules in this study are represented by only a single conformer.
In comparison, the commercial software COSMOtherm reports a similar uncertainty of \SIrange{0.40}{0.45}{\kilo\cal \per \mol}~\cite{Letcher2007, Klamt2015, Vermeire2021} for predicting the solvation free energy of neutral solutes, though it uses an ensemble of conformers.
However, the comparison may not be entirely fair, as the parameters of openCOSMO-RS 24a are directly adjusted to the data used in this study, whereas the accuracy reported by was likely based on a larger dataset.

Table \ref{tbl:comparison} presents a performance comparison of openCOSMO-RS 24a, COSMOtherm 24 TZVP, and COSMOtherm 24 FINE.
All calculations are conducted using only the lowest energy conformer, providing a fair comparison since openCOSMO-RS 24a currently lacks the capability to integrate multiple conformers.
Additional calculations using multiple conformers for TZVP and FINE parameterizations are included in the appendix.
Nevertheless, for the dataset examined in this study, incorporating multiple conformers does not significantly impact the results.
The results are categorized by solute type regarding its hydrogen binding capabilities. There is no universally accepted definition of a hydrogen-bonding molecule in the context of COSMO-RS modeling. However, we classify molecules based on the presence of areas in the $\sigma$-profile with an absolute screening charge density larger than the $\sigma_{\mathrm{HB}}$ threshold.
This method allows for a consistent categorization of molecule types, which directly relates to the terms in the interaction equations.

\begin{table}[h]
  \centering
  \caption{Comparison of openCOSMO-RS 24a, COSMOtherm 24 TZVP, and COSMOtherm FINE for infinite dilution activity coefficients, partition coefficients, and solvation free energies. The calculations for all three models were performed only with the lowest energy conformer.}
  \label{tbl:comparison}
  \resizebox{\textwidth}{!}{%
  \begin{tabular}{l c c c c}
    \toprule
    & & \textbf{\makecell{openCOSMO-RS\\ 24a}} & \textbf{\makecell{COSMOtherm\\ 24 TZVP}} & \textbf{\makecell{COSMOtherm\\ 24 FINE}} \\
    \midrule
    \textbf{IDAC} [-] & \textbf{$N_{\mathrm{datapoints}}$} & \textbf{AAD} & \textbf{AAD} & \textbf{AAD} \\
    \midrule
    non-HB & 568 & 0.41 & 0.43 & 0.40 \\
    HB acceptor & 172 & 0.63 & 0.53 & 0.35 \\
    HB donor & 35 & 0.64 & 0.93 & 0.40 \\
    HB acceptor/donor & 107 & 0.78 & 0.66 & 0.33 \\
    \textbf{Total} & \textbf{882} & \textbf{0.51} & \textbf{0.50} & \textbf{0.38} \\
    \midrule
    \textbf{Partition coefficients} [-] & & & & \\
    \midrule
    non-HB & 68 & 0.36 & 0.54 & 0.46 \\
    HB acceptor & 104 & 0.84 & 0.63 & 0.50 \\
    HB donor & 12 & 0.25 & 0.28 & 0.23 \\
    HB acceptor/donor & 112 & 0.99 & 0.76 & 0.59 \\
    \textbf{Total} & \textbf{296} & \textbf{0.76} & \textbf{0.64} & \textbf{0.51} \\
    \midrule
    \textbf{Solvation free energies} [\SI{}{\kilo\cal\per\mol}] & & & & \\
    \midrule
    non-HB & 434 & 0.36 & 0.34 & 0.32 \\
    HB acceptor & 775 & 0.40 & 0.47 & 0.45 \\
    HB donor & 69 & 0.52 & 0.39 & 0.32 \\
    HB acceptor/donor & 851 & 0.54 & 0.53 & 0.40 \\
    \textbf{Total} & \textbf{2129} & \textbf{0.45} & \textbf{0.46} & \textbf{0.40} \\
    \midrule
    \textbf{Overall} & \textbf{3307} & \textbf{0.49} & \textbf{0.50} & \textbf{0.41} \\
    \bottomrule
  \end{tabular}
  }
\end{table}

Overall, for \ac{IDAC} and solvation free energies, in the observed dataset, openCOSMO-RS 24a performs comparable to COSMOtherm 24 TZVP. Only for partition coefficients COSMOtherm 24 TZVP is more accurate. The COSMOtherm 24 FINE model delivers more accurate results for the majority of the systems in the dataset. 
This is a first step to improving openCOSMO-RS in a more general fashion for neutral molecules with current work focusing on improvements like the combinatorial term, temperature dependency, dispersion interactions, multiple conformers and polarizability effects.


\section{Conclusions}
In this work, we extended openCOSMO-RS to be able to calculate solvation free energies.
To do so, we developed a \ac{QSPR} model to predict the molar volumes of the solvents required for calculating solvation free energies.
By modifying the openCOSMO-RS conformer workflow and combining in total 3307 data points of activity coefficients, solvation free energies, and partition coefficients, we parameterized openCOSMO-RS 24a based on ORCA 6.0.

The openCOSMO-RS 24a parameterization based on ORCA 6.0 achieved an \ac{AAD} of \SI{0.45}{\kilo\cal \per \mol} for predicting solvation free energies, which is comparable to the uncertainty of \SI{0.45}{\kilo\cal \per \mol} that is reported by the commercial software COSMOtherm.
For predicting the partition coefficients, openCOSMO-RS 24a achieves an \ac{AAD} of \SI{0.76}{}.
Furthermore, openCOSMO-RS 24a showed an improvement in predicting activity coefficients with an \ac{AAD} of \SI{0.51}{} compared to an \ac{AAD} of \SI{0.65}{} with the previous openCOSMO-RS 22 parameterization.

While the openCOSMO-RS 24a parameterization performs well when representing each molecule with a single conformer, future work will focus on extending the model to handle conformer ensembles.
Additionally, we plan to extend openCOSMO-RS to ionic solutes, as current models often struggle to accurately predict the solvation free energies of ionic compounds\cite{Kroeger2020, Nevolianis2023c, Zheng2023}.

The performance of openCOSMO-RS in predicting liquid phase properties, even with a single conformer, demonstrates that the inclusion of additional chemical descriptors to surface charge improves the model's accuracy.
Moreover, substantial efforts have been made to integrate openCOSMO-RS into ORCA 6.0, enabling users to directly access a variety of liquid phase properties; a capability that was previously unavailable.
This integration marks a significant advancement, providing users with a powerful tool for comprehensive property prediction within a single software environment.




\bibliographystyle{elsarticle-harv} 
\bibliography{literature}

\appendix
\label{appendi}
\newpage

\section{Performance of COSMOtherm 24 TZVP and COSMOtherm 24 FINE using multiple conformers}

\begin{table}[H]
  \centering
  \caption{Performance of COSMOtherm 24 TZVP and COSMOtherm 24 FINE using multiple conformers for predicting infinite dilution activity coefficients, partition coefficients and solvation free energies.}
  \label{tbl:comparison_mc}
  \resizebox{\textwidth}{!}{%
  \begin{tabular}{l c c c}
    \toprule
    & & \textbf{COSMOtherm 24 TZVP MC} & \textbf{COSMOtherm 24 FINE MC} \\
    \midrule
    \textbf{IDAC [-]} & \textbf{Count} & \textbf{AAD} & \textbf{AAD} \\
    \midrule
    non-HB & 568 & 0.43 & 0.40 \\
    HB acceptor & 172 & 0.54 & 0.36 \\
    HB donor & 35 & 0.94 & 0.41 \\
    HB acceptor/donor & 107 & 0.65 & 0.27 \\
    \textbf{Total} & \textbf{882} & \textbf{0.50} & \textbf{0.38} \\
    \midrule
    \textbf{Partition coefficients [-]} & & & \\
    \midrule
    non-HB & 68 & 0.56 & 0.46 \\
    HB acceptor & 104 & 0.63 & 0.52 \\
    HB donor & 12 & 0.28 & 0.22 \\
    HB acceptor/donor & 112 & 0.78 & 0.59 \\
    \textbf{Total} & \textbf{296} & \textbf{0.66} & \textbf{0.52} \\
    \midrule
    \textbf{Solvation free energies [$\mathrm{kcal\ mol^{-1}}$]} & & & \\
    \midrule
    non-HB & 434 & 0.34 & 0.32 \\
    HB acceptor & 775 & 0.46 & 0.45 \\
    HB donor & 69 & 0.37 & 0.31 \\
    HB acceptor/donor & 851 & 0.52 & 0.39 \\
    \textbf{Total} & \textbf{2129} & \textbf{0.46} & \textbf{0.40} \\
    \midrule
    \textbf{Overall} & \textbf{3307} & \textbf{0.51} & \textbf{0.42} \\
    \bottomrule
  \end{tabular}
  }
\end{table}

\end{document}